\documentclass[smallextended]{svjour3} 
\smartqed 
\usepackage{graphicx}
\usepackage{subcaption}
\usepackage{amsmath}
\usepackage{amssymb}
\usepackage{natbib}
\usepackage[colorlinks=true,citecolor=blue,linkcolor=black]{hyperref}
\usepackage{slashbox}
\usepackage{lipsum}
\usepackage{multirow}
\usepackage{multicol}
\usepackage{url}
\usepackage{adjustbox}
\usepackage{rotating}
\usepackage{amsmath}

\begin{document}


\title{Multimodel  Response Assessment for Monthly Rainfall Distribution in Some Selected Indian Cities Using Best Fit Probability as a Tool}
\titlerunning{Rainfall probability distribution for twenty Indian cities}

%
%
%

\author{Anumandla Sukrutha \and 
Sristi Ram Dyuthi \and 
Shantanu Desai}
\institute{Anumandla Sukrutha \at  Dept. of Electrical Engineering, IIT Hyderabad, Kandi, Telangana 502285, India \email{ee14btech11002@iith.ac.in}
\and Sristi Ram Dyuthi \at Dept. of Electrical Engineering, IIT Hyderabad, Kandi, Telangana 502285, India \email{ee14btech11031@iith.ac.in}
\and Shantanu Desai \at Dept. of Physics, IIT Hyderabad, Kandi, Telangana 502285, India \email{shantanud@iith.ac.in}}

\authorrunning{A. Sukrutha et al}
\date{Received: date / Accepted: date}
\maketitle
\begin{abstract}
We carry out  a study of the statistical distribution of rainfall precipitation data for 20 cites in India.  We have  determined the best-fit probability distribution for  these cities from the monthly precipitation data spanning 100 years of observations from 1901 to 2002. To fit the observed data, we considered 10 different distributions. The efficacy of the fits for these distributions was evaluated using four empirical non-parametric  goodness-of-fit tests namely Kolmogorov-Smirnov, Anderson-Darling, Chi-Square,  Akaike information criterion, and Bayesian Information criterion. Finally, the best-fit distribution using each of these tests were reported,  by combining the results from the model comparison tests. We then find that for most of the cities, Generalized Extreme-Value Distribution or Inverse Gaussian Distribution most adequately fits the observed data.
\end{abstract}


\section{Introduction}
%
%

%
%
%
%
Establishing a probability distribution that provides a good fit to the monthly average precipitation  has long  been a topic of  interest in the fields of hydrology, meteorology, agriculture~\citep{Fisher}. The knowledge  of  precipitation at a given location  is  an  important  prerequisite  for agricultural planning and management. Rainfall is the main source of precipitation. Studies of  precipitation provide  invaluable knowledge about rainfall statistics. For rain-fed agriculture, rainfall is the single most important  agro-meteorological  variable  influencing  crop  production~\citep{Wallace,rockstrom}.  
In  the  absence  of  reliable physically  based  seasonal  forecasts,  crop  management  decisions  and  planning  have  to  rely  on statistical  assessment  based  on  the  analysis  of  historical  precipitation  records. It has been shown by \cite{Fisher}, that the statistical distribution of rainfall is more important than the total amount of rainfall for  the yield of crops.
Therefore, detailed statistical studies of rainfall data for a variety of countries   have been carried out for more than 70 years along with fits to multiple probability distribution~\citep{ghosh,Sharma,Tao}. 
We recap some of these studies for stations, both in India, as well as those outside India.
 
\cite{Mooley} first carried out a detailed statistical analysis of the rainfall distribution during southwest and northeast monsoon seasons at selected stations  in India with deficient rainfall, and found that the Gamma distribution provides the best fit. 
~\citet{Stephenson} showed that the outliers in the rainfall distribution for the summers of 1986 to 1989 throughout India can be well fitted  by the gamma and Weibull distributions.~\cite{Deka} found that the logistic distribution is the optimum distribution for the annual rainfall distribution for  seven districts in north-East India.
~\cite{Sharma} found based on daily rainfall data for Pantnagar spanning 37 years, that the lognormal and gamma distribution provide the best fit probability distribution for  the annual and monsoon months, whereas the Generalized extreme value provides the best fit after considering only the weekly data. Most recently, \cite{Kumar} analyzed the statistical distribution of rainfall in Uttarakhand, India and found  that the Weibull distribution performed the best.  However, one caveat with some of the above studies is that only a handful of  distributions were considered for fitting the rainfall data, and sometimes
 no detailed model comparison tests  were done to find the most adequate  distribution.

A large number of statistical studies have similarly been done for rainfall precipitation data for stations outside India. For brevity, we only mention a few selected studies to illustrate the diversity in the best-fit distribution found from these studoes. In Costa Rica, normal distribution provided the best fit to the annual rainfall distribution~\citep{Waylen}.  
A generalized extreme value distribution has been used for Louisiana~\citep{naghavi}.   Gamma  distribution provided the best fit for rainfall data in Saudi Arabia~\citep{abdullah}, Sudan~\citep{mohamed2015} and Libya~\citep{ejadid}. ~\cite{Mahdavi} studied the rainfall statistics for 65 stations in the Mazandaran and Golestan provinces in Iran and found that the Pearson and log-Pearson distribution provide the best fits to the data.
~\cite{Choi} found that Gumbel distribution provides the most reasonable fit to the data in South Korea.~\citet{ghosh} found that the extreme value distribution provides the best fit to the Chittagong monthly rainfall data during the rainy season, whereas for Dhaka, the gamma distribution provides a better fit.

Therefore, we can see from these whole slew of studies,  that no single  distribution can accurately describe the rainfall distribution.
The selection depends on the characteristics of available rainfall data as well as the statistical tools used for model selection.

The main objective of the current study is to complement the above studies and to determine the best fit probability distribution for the monthly average precipitation data of 20 selected stations throughout India, using multiple goodness of fit tests.


\section{Datasets and Methodology}
\label{sec:dataset}
\begin{figure}[h]
\centering
       \includegraphics{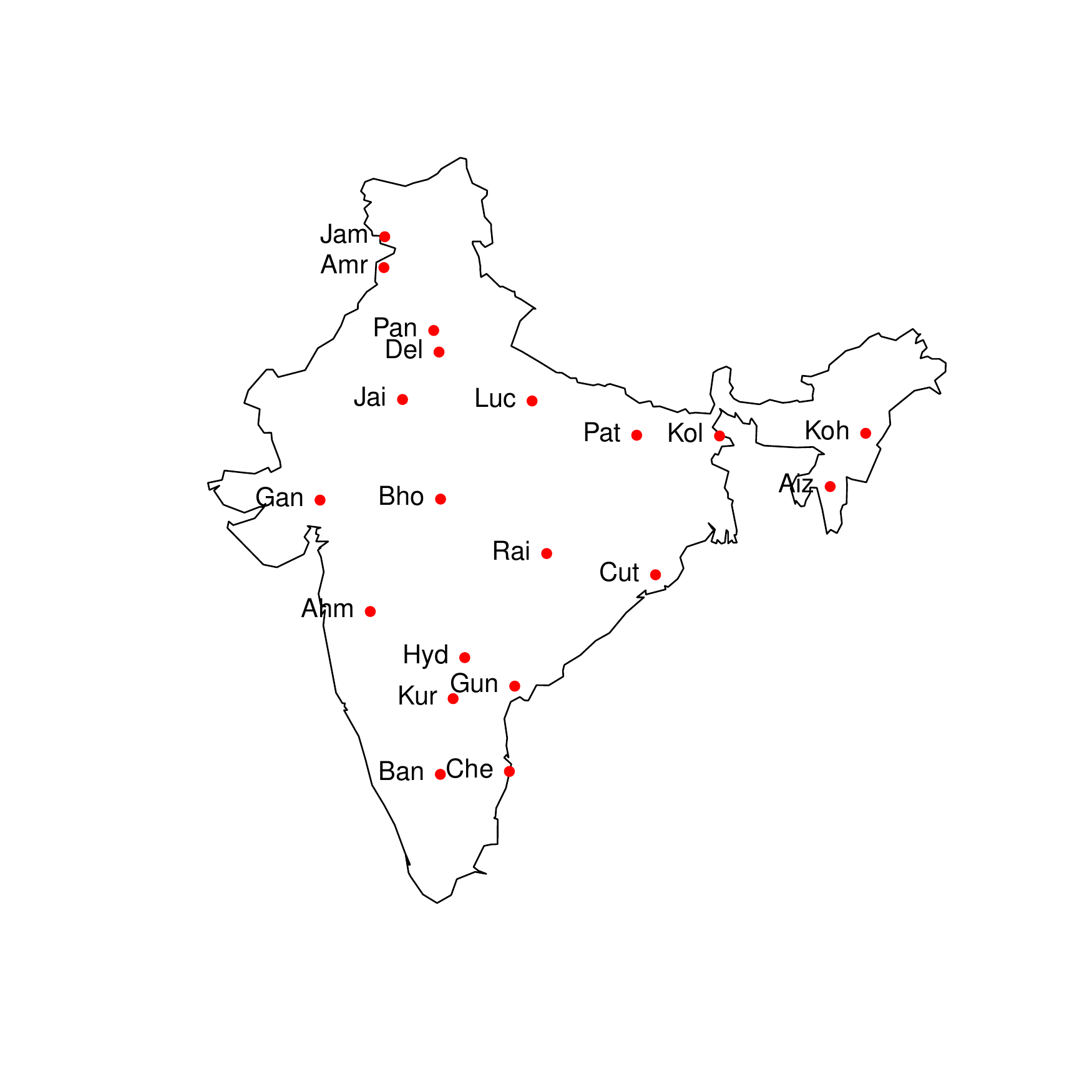}
       \caption{Map showing location of various stations throughout India for which rainfall statistics and best-fit distributions were obtained. Each red point represents a station and next to it we show its  first three letters. The full names of the cities can be found in Table~\ref{tab:1}. This plot has been made with the {\tt ggplot}\citep{Rplot} data visualization package in  the {\tt R} programming language, where ``gg'' in {\tt ggplot} is an abbreviation for "Grammar of graphics".}
\label{fig:map}       
\end{figure}

The datasets employed here for our study span  a 100-year period from 1901 to 2002, and is based on  records  collected by the Indian Meteorological Department. This  data  can be downloaded from \url{http://www.indiawaterportal.org/met_data/}.
From these, we selected 20 stations, covering the breadth of the country for our study. 
The stations used for this study are Gandhinagar, Guntur, Hyderabad, Jaipur, Kohima, Kurnool, Patna, Aizwal, Bhopal, Ahmednagar, Cuttack, Chennai, Bangalore, Amritsar, Guntur, Lucknow, Kurnool, Jammu, Delhi, and Panipat. The location of these stations on a map of India is shown in Fig.~\ref{fig:map}. Detailed rainfall statistics
for each of these stations can be found in Table~\ref{tab:1}.

\begin{figure}[h]
\centering
       \includegraphics[scale = 0.3]{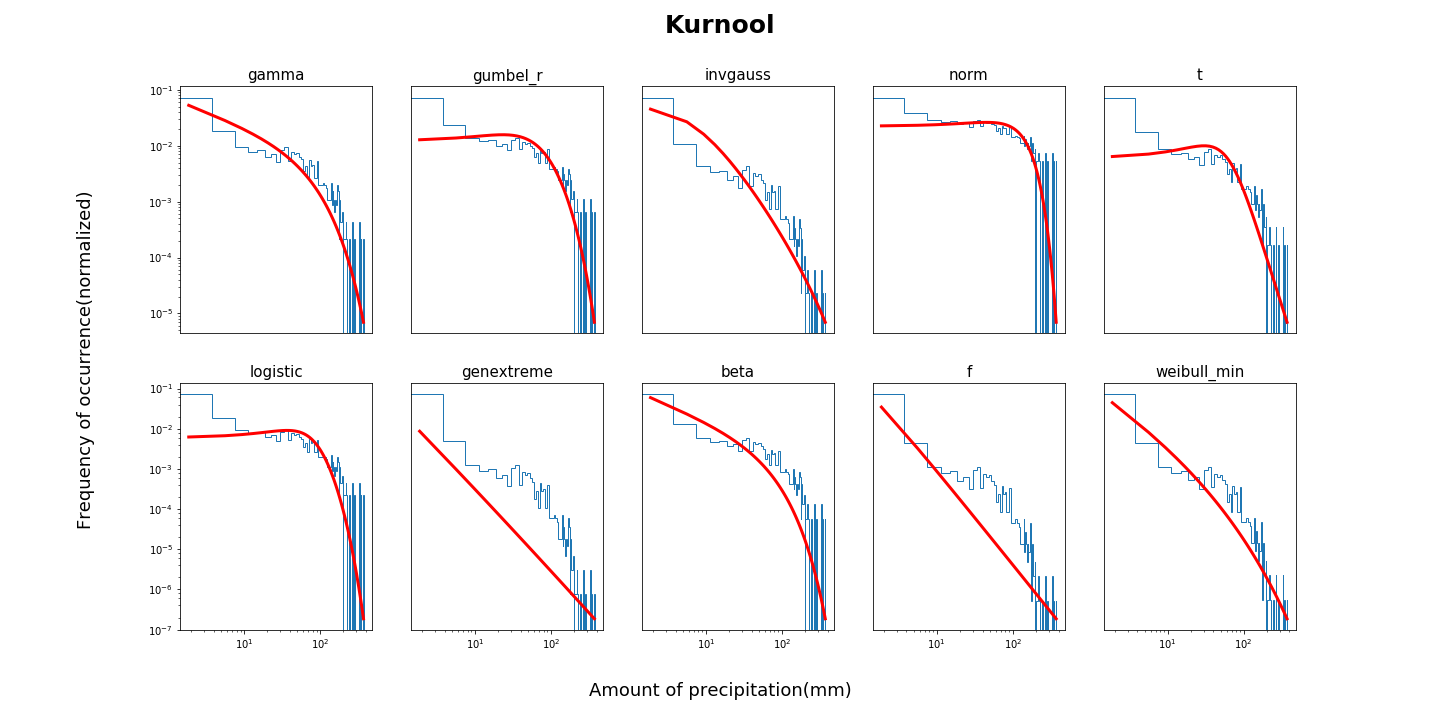}
       \caption{Histogram of the  monthly precipitate data at Kurnool (blue lines) along with best fit for each of the 10 probability distributions functions considered.}
\end{figure}
\begin{figure}[h]
\centering
       \includegraphics[scale = 0.3]{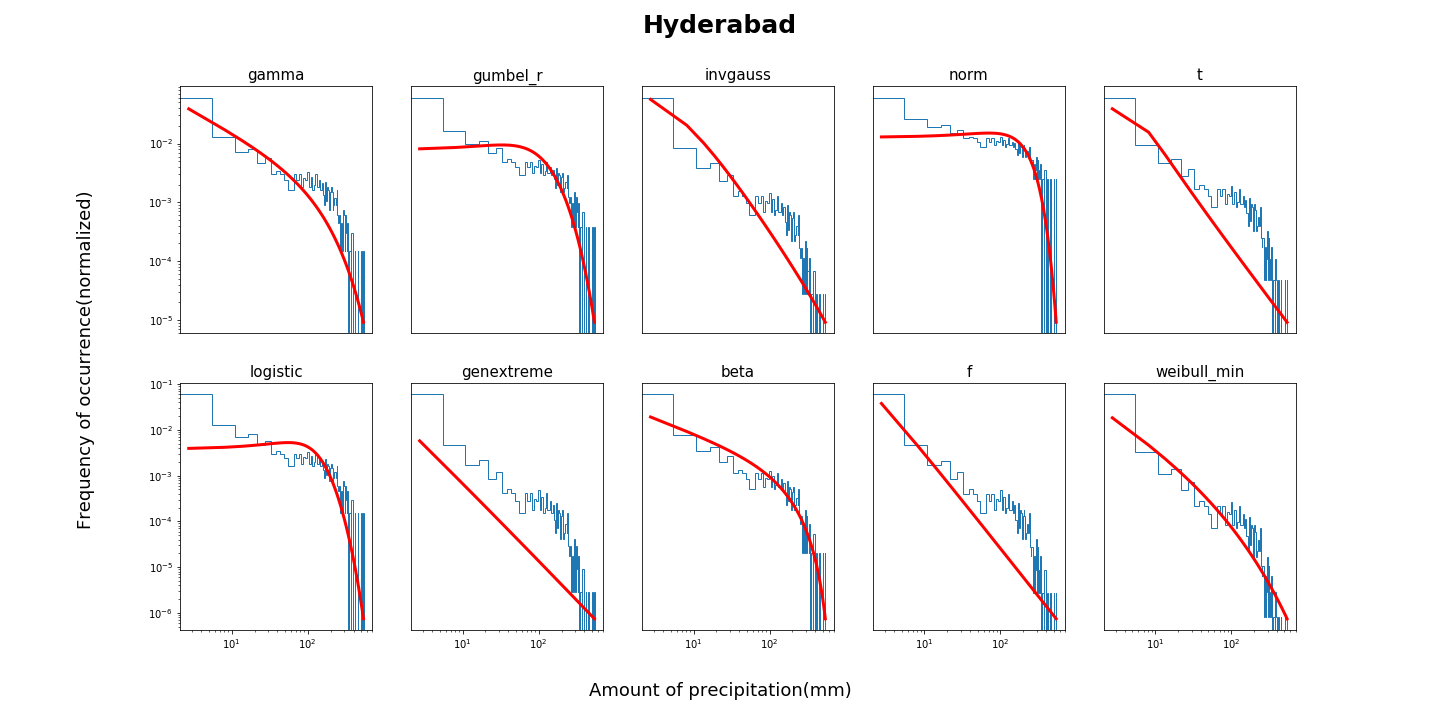}
       \caption{Histogram of  the monthly precipitate data at Hyderabad (blue lines) along with best fit for each of the 10 probability distributions functions considered.}
\end{figure}
\begin{figure}[h]
\centering
       \includegraphics[scale = 0.3]{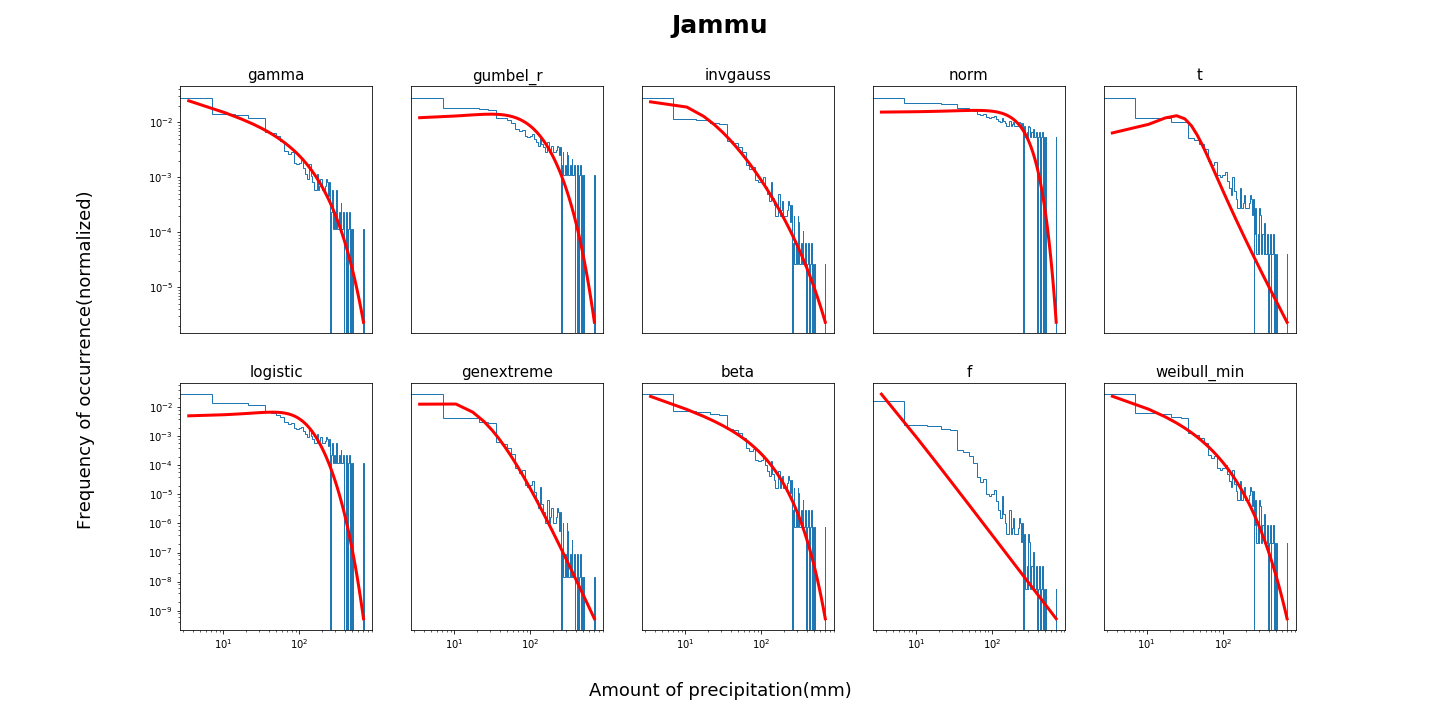}
       \caption{Histogram of the  monthly precipitate data at Jammu (blue lines) along with best fit for each of the 10 probability distributions functions considered.}
\end{figure}
\begin{figure*}[h]
\centering
       \includegraphics[scale = 0.3]{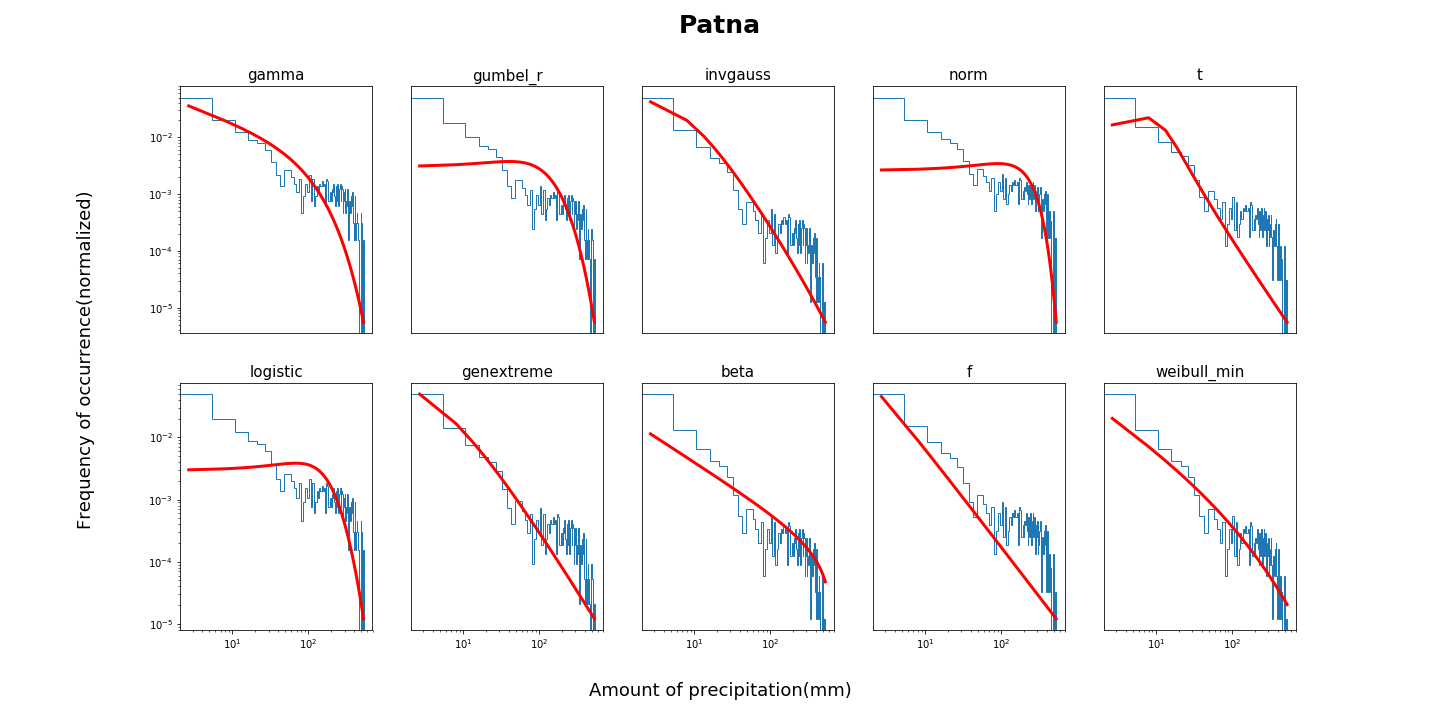}
       \caption{Histogram of  the monthly precipitate data at Patna (blue lines) along with best fit for each of the 10 probability distributions functions considered.}
\end{figure*}
The list of probability distributions considered for fitting the rainfall data  include: Gamma, Fisher, Inverse Gaussian, Normal, Student$-$t, LogNormal, Generalized Extreme value, Weibull, and  Beta distributions.
The mathematical expressions for the probability density functions of these distributions can be found in  Table~\ref{dist}, and have been adapted from \citet{astroML,ghosh}. All of these  distributions have  been previously used for similar studies (eg.~\citet{ghosh,Sharma} and the other references listed in the introductory section).  For each station, we find the best-fit parameters for each of these probability distribution using maximum-likelihood analysis. To select the best-fit distribution for a given station, we then use multiple model comparison techniques  to rank each distribution for every city. We now describe the model comparison techniques used.

\begin{table*}[h!]
\centering
\caption{Probability density functions of different distributions used to fit the rainfall data~\citep{astroML,ghosh}.}
\label{dist}
\begin{tabular}{|c|c|}\hline
Distribution & Probability density function \\ \hline
Normal & $f(x) = \frac{1}{\sqrt{2\pi{\sigma}^2}}\exp{-\frac{(x-\mu)^2}{2{\sigma}^2}}$  \\ \hline
Lognormal & $f(x) = \frac{1}{x\sqrt{2\pi{\sigma}^2}}\exp{-\frac{(\ln x-\mu)^2}{2{\sigma}^2}}$  \\\hline
Gamma & $f(x) = \frac{1}{\theta_k}\frac{x^{k-1}\exp(-x/\theta)}{\Gamma(k)}$ \\\hline
Inverse Gaussian & $f(x) = {\frac{\lambda}{2\pi x^3}}^{0.5}\exp{\frac{-\lambda(x-\mu)^2}{2\mu^2x}}$\\\hline
GEV  & $f(x) = \frac{1}{\sigma}[1-k\frac{x-\mu}{\sigma}]^{1/k-1}{\exp[-(1-k\frac{x-\mu}{\sigma})]}^{1/k}$  \\\hline
Gumbel & $f(x) = {1/\beta} \exp(-z+\exp(-z))),z=\frac{x-\mu}{\beta}$  \\\hline
Student-t & $f(x) = \frac{\Gamma(\frac{v+1}{2})}{\sqrt{v\pi}\Gamma(v/2)}{(1+\frac{x^2}{v})}^{\frac{-v-1}{2}}$ \\\hline
Beta & $f(x) = \frac{x^{\alpha -1}(1-x)^{\beta -1}}{\frac{\Gamma(\alpha)\Gamma(\beta)}{\Gamma(\alpha+\beta)}}$  \\\hline
Weibull&$f(x) = \frac{k}{\lambda}(\frac{x}{\lambda})^{k-1}e^{-(\frac{x}{\lambda})^{k}} $ 
\\\hline
Fisher&$\frac{\sqrt[]{\frac{(d_{1}x)^{d_{1}}d_{2}^{d_{2}}}{(d_{1}x+d_{2})^{d_{1}+d_{2}}}}}{xB(\frac{d_{1}}{2},\frac{d_{2}}{2})}$\\\hline

\end{tabular}

\end{table*}

\subsection{Model Comparison tests}
We use multiple model comparison methods
 to  carry out hypothesis testing and select  the best distribution for the precipitation data.
For this purpose, the goodness of fit   tests used  include non-parametric distribution-free tests such as   Kolmogorov$-$Smirnov Test, Anderson-Darling Test, Chi$-$square test, and information-criterion tests such as Akaike and Bayesian Information Criterion. For each of the probability distributions, we find the best-fit 
parameters for each of the stations using least-squares fitting and then carry out each of these tests. We now describe these tests.
\subsubsection{Kolmogorov-Smirnov Test}
The Kolmogorov-Smirnov  (K-S) test~\citep{astroML}   is a non-parametric test  used to decide if a sample is selected  from a population with a specific distribution. The K-S test compares  the empirical distribution function (ECDF) of two samples. Given $N$ ordered data points $y_1, y_2, ..., y_N,$ the ECDF is defined as 
\begin{equation}
E_N = n(i)/N,
\end{equation}
where $n(i)$ indicates  the total  number of points less than $y_i$, after sorting the $y_i$ in increasing order. This is a step function, whose value  increases by $1/N$ for  each sorted data point.\par
The K-S test is based on the maximum distance (or supremum) between the empirical distribution function and the normal cumulative distributive function. An attractive feature of this test is that the distribution of the K-S test statistic itself does not depend on the statistics of the parent distribution from which the samples are drawn. Some limitations are that it  applies only to continuous distributions and tends to be more sensitive near the center of the distribution than at the tails.\par
The Kolmogorov-Smirnov test statistic is defined as: 

\begin{equation}
	 \underset{1\leqslant i \leqslant N}{\text{max}}
	 (F(y_i)-\frac{i-1}{N},\frac{i}{N}-F(y_i)),  \\
\end{equation}

\noindent where $F$ is the cumulative distribution function of the samples  being tested.
If the probability that a given value of $D$ is very small (less than a certain critical value, which can be obtained from tables)  we can reject the null hypothesis that the two samples are drawn from the same underlying distributions at a given confidence level.

\subsubsection{Anderson-Darling Test}
The Anderson-Darling test~\citep{astroML} is another test (similar to K-S test), which  can evaluate whether a sample of data came from a population with a specific distribution. It is a modification of the  K-S test, and gives more weight to the  tails compared to  the K-S test.  Unlike the K-S test, the Anderson-Darling test makes use of the specific distribution in calculating the critical values. This has the advantage of allowing a more sensitive 
test. However, one disadvantage  is that  the critical values must be calculated separately  for each distribution. The Anderson-Darling test statistic is defined as follows~\citep{astroML}:

 \begin{equation} \label{eq1}
 \begin{split}
 A^2 & = -N - \sum\limits_{i=1}^{N}\frac{(2i-1)}{N}[\log F(y_i) \\
  &  + \log(1-F(y_{N+1-i}))]
 \end{split}
 \end{equation}

\noindent where $F$ is the cumulative distribution function of the specified distribution and  $y_i$ denotes the sorted data. The test is a one-sided test and the hypothesis that the data is sampled from a specific distribution is rejected if the test statistic, $A$, is greater than the critical value. For a given distribution, the Anderson-Darling statistic may be multiplied by a constant (depending on the sample size, $n$). These constants have been tabulated by ~\cite{Stephens}.
\subsubsection{Chi-Square Test}
The chi-square test~\citep{Cochran} is used to test if a sample of data is obtained from a population with a specific distribution. An attractive feature of the chi-square goodness-of-fit test is that it can be applied to any univariate distribution for which you can calculate the cumulative distribution function. The chi-square goodness-of-fit test is usually applied to binned data.   The chi-square goodness-of-fit test can be applied to discrete distributions such as the binomial and the Poisson distributions. The Kolmogorov-Smirnov and Anderson-Darling tests can only be applied  to continuous distributions. For the chi-square goodness-of-fit computation, the data are sub-divided into $k$ bins and the test statistic is defined as follows:
\begin{equation}
{\chi}^2=\sum_{i=1}^{k} \frac{(O_i - E_i)^2}{E_i},
\end{equation} 
\noindent where $O_i$ is the observed frequency for bin $i$ and $E_i$ is the expected frequency for bin $i$. The expected frequency is calculated by 

\begin{equation}
E_i = N(F(Y_u)-F(Y_l)),
\end{equation}
\noindent where $F$ is the cumulative distribution function for the distribution being tested, $Y_u$ is the upper limit for class $i$, $Y_l$ is the lower limit for class $i$, and $N$ is the sample size.

This test is sensitive to the choice of bins. There is no optimal choice for the bin width (since the optimal bin width depends on the distribution). For our analysis,  since there were a total of 1224 data points, we have chosen 100 bins, so that there were sufficient data points in each bin.
For the chi-square approximation to be valid, the expected frequency of events in each bin should be at least five.  The test statistic follows, approximately, a chi-square distribution with ($k - c$) degrees of freedom, where $k$ is the number of non-empty cells, and $c$  is  the number of estimated parameters (including location, scale,  and shape parameters) for the distribution + 1. Therefore, the hypothesis that the data are from a population with the specified distribution is rejected if: 
\begin{equation}
{\chi}^2 \geqslant {\chi_{1-\alpha,k-c}}^2,
\end{equation}
where ${\chi_{1-\alpha,k-c}}^2$ is the chi-square critical value with $k - c$ degrees of freedom and significance level $\alpha$. 
\subsubsection{AIC and BIC}
The Akaike Information Criterion (AIC)~\citep{Liddle,Kulkarni} is a way of selecting a model from an input set of models. It can be derived by an approximate minimization of  the Kullback-Leibler distance between the model and the truth. It is based on information theory, but a heuristic way to think about it is as a criterion that seeks a model, which has a good fit to the truth with very  few parameters. 

It is defined as ~\citep{Liddle}:
\begin{equation}
\rm{AIC} = -2\log(\mathcal{L}) + 2K
\end{equation}
where $\mathcal{L}$ is the likelihood which denotes the probability of the data given a model, and $K$ is the number of free parameters in the model. AIC scores are often shown as $\Delta$AIC scores, or difference between the best model (smallest AIC) and each model (so the best model has a $\Delta$AIC of zero).\par
The bias-corrected information criterion, often called AICc, takes into account the finite  sample size, by essentially increasing the relative penalty for model complexity with small data sets. It is defined as~\citep{Kulkarni}:
\begin{equation}
\rm{AICc} = -2\log(\mathcal{L})+2 \frac{K(K+1)}{N-K-1}
\end{equation}
\noindent where $\mathcal{L}$ is the likelihood and  $N$ is the sample size. 
For this study we have used AICc for evaluating model efficiacy.

Bayesian information criterion (BIC) is also an alternative  way of selecting a model from a set of models.  It is an approximation to Bayes factor between two models. It is defined as~\citep{Liddle}: 
\begin{equation}
\rm{BIC} = -2\log(\mathcal{L})+K\log(N)
\end{equation}
When comparing the BIC values for two models, the model with the smaller BIC value is considered better. In general, BIC penalizes models with more parameters more than AICc does.

\section{Results and Discussion}
The summary statistics for the amount of monthly precipitation data for the above mentioned stations are summarized in Table~\ref{tab:1}, where the minimum, maximum, mean, standard deviation (SD), coefficient of variation (CV), skewness, and kurtosis are shown.
The monthly rainfall dataset indicates that the monthly rainfall was strongly positively skewed for Gandhinagar, Jaipur, Amritsar, Delhi, and Panipat stations. Aizwal, Kohima, and Cuttack show negative values of kurtosis.
The distributions listed above are fitted for  each of the selected locations. For 
brevity in this manuscript,  we show the plots  for only four cities. These can be found in Figures [2-5], which  illustrate the fitted distribution for Kurnool, Hyderabad, Jammu, and Patna. These plots are mainly for illustrative purposes. More detailed information about the rainfall distribution can be gleaned from the statistical fits to different distributions.
Similar plots for the remaining stations have been uploaded on a google drive, whose link is provided at the end of this manuscript.\\

\begin{table}[h]
\caption{Summary statistics of monthly precipitate data for the selected stations during the years (1901-2002). We note that all units of dimensional quantities are in mm.}
\begin{tabular}{|c|c|c|c|c|c|c|c|}\hline
& Min. & Max. & Mean  & Standard  & Coeff  & Coeff. & Kurtosis \\
& & & & Deviation & of variation & of skewness & \\ \hline
Kohima & 0 & 802.43 & 196.33 & 177.67 & 0.91 & 0.77 & -0.24 \\\hline
Jaipur & 0 & 517.61 & 48.6 & 83.53 & 1.72 & 2.28 & 5.26 \\ \hline
Kolkata  & 0 &  892.15 &  132.15 &  148.63 &  1.13 &  1.31 &  1.474 \\ \hline
Raipur & 0  & 635.98 & 105.38 &  140.33 &  1.33 &  1.33 &  0.72 \\ \hline
Gandhinagar &  0 &  694.2 &  56.42 &  105.18 &  1.86 &  2.33 &  5.36 \\ \hline
Hyderabad & 0 &  544.26 &  70.06 &  89.41 &  1.28 & 1.53 &  2.19 \\ \hline
Aizawl & 0 &  1065.92 & 227.2 &  221.48 &  0.98 &  0.8 &  -0.311 \\ \hline
Bhopal & 0  & 725.72 &  89.53 &  140.91 &  1.57 &  1.73 &  2.18 \\ \hline
Ahmednagar & 0 & 611.13 &  70.73 &  96.63 & 1.37 & 1.58 &  2.33 \\ \hline
Cuttack &  0 & 506.19  & 106.32 &  115.32 &  1.09 & 0.91 &  -0.34 \\ \hline
Chennai &  0 & 768.91 & 96.89 & 118.27 &  1.22 &  1.99 &  4.82 \\ \hline
Bangalore &  0  & 360.95 & 69.89 &  68.66 &  0.98 &  1.08 &  0.78 \\ \hline
Patna & 0  & 534.69 &  90.96 &  121.9 &  1.34 &  1.39 & 0.9 \\ \hline
Amritsar & 0 & 416.06 &  39.16 & 59.15 &  1.51 &  2.61 &  8.02 \\ \hline
Guntur  & 0  & 438.45 &  65.66 &  74.58 & 1.14 & 1.44 &  2.24 \\ \hline
Lucknow & 0 & 619.08 & 74.85 & 113.6 & 1.52 & 1.76 & 2.43 \\ \hline
Kurnool  & 0 &  374.53 &  45.19 &  53.93 &  1.19 & 1.85 &  4.69 \\ \hline
Jammu &  0 &  704.43  &  60.88 &  83.41 &  1.37 &  2.59 &  8.35 \\ \hline
Delhi &  0  & 511.54 &  47.45 &  80.67 &  1.7 &  2.47 &  6.58 \\ \hline 
Panipat &  0  & 463.83 &  43.58 &  69.103 &  1.59 &  2.33 &  5.87 \\ \hline 
\end{tabular}
\label{tab:1}
\end{table}

The test statistics for K-S test (D), Anderson-Darling Test ($A^{2}$), Chi-square test ($\chi^{2}$), AICc, and  BIC were computed for the  ten probability distributions. The AICc and BIC values for each of these 10 distributions and 20 cities can be found on the  google drive, which documents this analysis. The probability distribution that   fits a given data the best (using the largest $p$-value) according to each of the above criterion  is shown in Table~\ref{stationwide}.

\begin{table}[h]
\centering
\caption{Station-wise best ranked probability distribution using different goodness of fit tests. F stands for the Fisher distribution, t stands for Students-t distribution and GEV stands for Generalized Extreme value distribution.}
\label{stationwide}
\begin{tabular}{|l|l|l|l|l|l|}\hline
Study Location & KS           & AD   & Chi Square   & AIC(c)          & BIC  \\\hline
Patna          & F            & F    & GEV   & F            & Beta \\\hline
Kurnool        & F            & F    & Weibull & F            & Beta \\\hline
Jaipur         & F            & F    & Inv. Gauss     & F            & Beta \\\hline
Chennai        & F            & F    & Gamma        & F            & Beta \\\hline
Hyderabad            & F            & F    & Inv Gauss     & F            & Beta \\\hline
Lucknow        & F            & F    & Inv. Gauss     & F            & Beta \\\hline
Bangalore      & F            & F    & Weibull & F            & Beta \\\hline
Kohima         & Weibull & Beta & Beta         & Weibull & Beta \\\hline
Aizawl        & Weibull & Beta & Gamma        & Weibull & Beta \\\hline
Guntur         & F            & F    & F            & F            & Beta \\\hline
Panipat        & F            & F    & GEV   & F            & Beta \\\hline
Amritsar       & F            & F    & Inv. Gauss     & F            & Beta \\\hline
Cuttack        & F            & F    & GEV   & F            & Beta \\\hline
Gandhinagar    & F            & F    & Beta         & GEV   & t    \\\hline
Ahmednagar    & F            & F    & Inv. Gauss     & t            & Beta \\\hline
Raipur         & F            & F    & GEV   & F            & Beta \\\hline
Jammu          & F            & F    & Weibull & F            & Beta \\\hline
Kolkata        & F            & F    & F            & F            & Beta \\\hline
Bhopal         & F            & F    & Inv. Gauss     & F            & Beta \\\hline
Delhi          & F            & F    & Inv. Gauss     & F            & Beta \\\hline
\end{tabular}
\end{table}

\par For each station, we ranked
all the probability distribution functions, using each of the four model comparison techniques in
decreasing order of its $p$-value.
The best fit distribution amongst these, for each city was found after summing these ranks, and choosing the function with the smallest cumulative rank. A similar technique was also used in ~\cite{Sharma} to find the best distribution, which fits the rainfall data using multiple model comparison techniques. The best fit distribution for each station using this ranking technique is shown in Table~\ref{bestfit}. 
After obtaining the best fit, similar to ~\cite{ghosh}, we then calculate the first four sample L-moments for each station. L-moments are linear combinations of expectations of order statistics and are reviewed extensively in ~\cite{Hosking}. They are more robust estimates  of the central moments than the conventional moments. The first four L-moments are analogous to mean, standard deviation, skewness and kurtosis. These L-moments are shown in Table~\ref{lmoment}.

\begin{table*}[tbp]
\centering
\caption{Station-wise best fit distribution obtained by summing  the ranks of each of the distributions from all the  model comparison tests considered in Table~\ref{stationwide}.}
\label{bestfit}
\begin{tabular}{|c|c|}\hline
Study location & Best Fit \\\hline
Kohima & genextreme \\\hline
Jaipur& invgauss \\\hline
Kolkata& genextreme \\\hline
Raipur& genextreme\\\hline
Gandhinagar& genextreme \\\hline
Hyderabad&invgauss  \\\hline
Aizawl& gamma \\\hline
Bhopal& invgauss \\\hline
Ahmednagar& invgauss \\\hline
Cuttack&  genextreme\\\hline
Chennai&invgauss \\\hline
Banglore& genextreme\\\hline
Patna& genextreme\\\hline
Amritsar&  invgauss\\\hline
Guntur& Gumbel  \\\hline
Lucknow&invgauss \\\hline
Kurnool& Gumbel  \\\hline
Jammu& invgauss \\\hline
Delhi& invgauss\\\hline
Panipat & genextreme \\\hline

\end{tabular}
\\
\end{table*}

\begin{table}[]
\centering
\caption{Parameters estimates using sample L-moments (mean (L1), variance (L2), skewness (L3) , kurtosis (L4)) of the best fitted distributions}
\label{lmoment}
\begin{tabular}{|l|l|l|l|l|l|}\hline
Study Location &Best-Fit& Mean (L1) & Variance (L2) & Skewness (L3) & Kurtosis (L4)\\\hline

Kohima      &GEV & 196.33 & 98.56  & 0.21 & 0.02 \\\hline
Jaipur      &Inv Gauss& 48.6   & 36.27  & 0.57 & 0.27 \\\hline
Kolkata     &GEV & 132.15 & 77.77  & 0.33 & 0.07 \\\hline
Raipur      &GEV & 105.38 & 70.01  & 0.43 & 0.09 \\\hline
Gandhinagar &GEV & 56.42  & 44.44  & 0.61 & 0.29 \\\hline
Hyderabad   &Inv Gauss & 70.06  & 45.1   & 0.4  & 0.1  \\\hline
Aizawl     &Gamma & 227.2  & 121.87 & 0.24 & 0.01 \\\hline
Bhopal      &Inv Gauss & 89.53  & 65.42  & 0.53 & 0.18 \\\hline
Ahmedanagar &Inv Gauss & 70.73  & 47.78  & 0.43 & 0.11 \\\hline
Cuttack     &GEV & 106.32 & 62.07  & 0.3  & 0.01 \\\hline
Chennai     &Inv Gauss & 96.89  & 58.01  & 0.39 & 0.17 \\\hline
Bangalore   &GEV & 69.89  & 37.14  & 0.26 & 0.07 \\\hline
Patna       &GEV & 90.96  & 60.58  & 0.44 & 0.11 \\\hline
Amritsar    &Inv Gauss & 39.16  & 26.01  & 0.51 & 0.27 \\\hline
Guntur      &Gumbel & 65.66  & 38.73  & 0.33 & 0.08 \\\hline
Lucknow     &Inv Gauss & 74.85  & 53.11  & 0.51 & 0.18 \\\hline
Kurnool     &Gumbel & 45.19  & 27.06  & 0.36 & 0.13 \\\hline
Jammu       &Inv Gauss & 60.88  & 37.41  & 0.48 & 0.26 \\\hline
Delhi       &Inv Gauss & 47.45  & 34.68  & 0.57 & 0.28 \\\hline
Panipat     &GEV & 43.58  & 30.62  & 0.54 & 0.25 \\\hline
\end{tabular}
\end{table}

Our results from each of the   model comparison tests are summarized as  follows:
\begin{itemize}
\item Using K-S  test (D), we find  that the Fisher distribution provides a good fit to the monthly precipitation data for all cities except  Kohima and Aizawl. For these cities, Weibull distribution provide the best fit.

\item Using Anderson-Darling Test ($A^{2}$), it is observed that the Fisher distribution is the best fit for all the cities except (again) for Kohima and Aizawl, for which the Beta distribution gives the best fit for both the cities.

\item Using Chi-square test ($\chi^{2}$), 
there is no one distribution which consistently provides 
the best fit for most of the   cities. Inverse Gaussian is the optimum fit for   seven cities, whereas Weibull and Generalized extreme  for three cities, Beta and Fisher for two cities each.  The locations of the corresponding cities can be found in Table~\ref{stationwide}.

\item Using  AICc, it is observed that the Fisher distribution provides best distribution for about 16 cities. 
The exceptions are again Kohima and Aizawl, for which Weibull is the most appropriate distribution.
Generalized extreme value distribution provides the best fit for Gandhinagar, whereas Students t-distribution provides the best fit for Ahmednagar.
 
\item For BIC,  we find that the beta distribution provides best distribution for all districts except Gandhinagar. Student-t distribution provides best fit for Gandhinagar.
\end{itemize}

If we then determine the best distribution from a combination of the above model comparison techniques using the ranking technique, we find  (cf. Table~\ref{bestfit}) that the generalized extreme value distribution is the most appropriate for eight cities, inverse Gaussian  for nine cities, Gumbel  for two cities, and gamma for one city. Therefore, although no one distribution provides the best fit for all stations, for most of them  can be best fitted using either the generalized extreme value or inverse Gaussian distribution.
\section{Implementation}
We have used the  python v2.7 environment. In addition, Numpy, pandas, matplotlib, scipy packages are used. Our codes to reproduce all these results can be found in \url{http://goo.gl/hjYn1S}. These can be easily applied to statistical studies of rainfall distribution for any other station.

\section{Comparison  to previous results}
A summary of some of the previous studies of rainfall distribution for various stations in India is outlined in the introductory section. An apples-to-apples comparison to these results is not straightforward, since they have not used the same model comparison techniques or considered all the 10 distributions which we have used. Moreover, the dataset and duration they have used is also different.  Nevertheless, we compare and contrast  the salient features of our conclusions with the previous results.

Among the previous studies, ~\cite{Sharma} have also found that Generalized extreme value distribution fits the weekly  rainfall data for Pantnagar. We also find that this distribution provides the best fit for eight cities.
The best-fit distribution which we found for Aizawl agrees with the results from ~\cite{Mooley,kuland,Bhaskar}. None of the previous studies have found the Inverse Gaussian or the Gumbel  distribution to be an adequate fit to the rainfall data. However, this could be because these two distributions were not fitted to the observed data in any of the previous studies. Inverse Gaussian and the Gumbel distribution have only recently been considered by ~\cite{ghosh} and \cite{Choi} for fitting the rainfall data in Bangladesh and Korea respectively.
We hope our results spur future studies to consider these distributions for fitting rainfall data in India.

\section{Conclusions}

We carried out a systematic study
to identify the best fit probability distribution for the monthly precipitation data at twenty selected stations distributed uniformly throughout India. The data showed that the monthly minimum and maximum precipitation at any time at any station ranged from 0 to 802 mm, which obviously indicates a large dynamic range. So identifying the best parametric distribution for the  monthly precipitation data could have a wide range of applications in agriculture, hydrology, engineering design, and climate research.

For each station, we fit the precipitation data  to 10 distributions described in Table~\ref{dist}. To determine the best fit among these distributions, we used five model comparison tests, such as K-S test, Anderson-Darling test, chi-square test, Akaike and Bayesian Information criterion. The results from these tests are summarized in Table 3.  For each model comparison test, we ranked each distribution according to its $p$-value and then added the ranks from all the four tests. The best-fit distribution for each city is the one with the minimum total rank, and  is tabulated  in Table~\ref{bestfit}. 
We find that no one distribution can adequately describe the rainfall data  for all the stations. For about nine cities,  the Inverse Gaussian distribution provides the best fit, whereas Generalized extreme value can adequately fit the rainfall distribution for about eight cities. Our study is the first one, which finds the Inverse Gaussian distribution to be the optimum fit for any station. Among the remaining cities, Gumbel and Gamma distribution are the best fit for two and one city respectively.

In the hope that this work would be of interest to researchers wanting to do similar analysis and to promote transparency in data analysis, we have made our analysis codes as well as data publicly available for anyone to reproduce this results as well as to do similar analysis on other rainfall datasets. This can be found at \url{http://goo.gl/hjYn1S}

\bibliography{rainfalldist}

\end{document}